\documentclass[reprint,aps,prb,amsmath,superscriptaddress,bibnotes,longbibliography]{revtex4-2}

\usepackage{mathptmx,newtxtext,newtxmath,xspace}
\usepackage{amsbsy,bm,bbold}
\usepackage{graphicx,color,xcolor}
\usepackage{fancyhdr}
\usepackage[colorlinks=true, 
linkcolor=blue, 
urlcolor=blue,
citecolor=blue]{hyperref}
\usepackage{psfrag}
\usepackage{enumitem}

\DeclareTextCompositeCommand{\r}{OT1}{A}{%
	\leavevmode\vbox{%
		\offinterlineskip
		\ialign{\hfil##\hfil\cr\char23\cr\noalign{\kern-1.15ex}A\cr}%
	}%
}

\begin{document}

\title{Interplay between electronic and lattice superstructures in La$_{2-x}$Ca$_{x}$CuO$_{4}$}

\widetext
\date{\today}
	
\author{S. Hameed}
\affiliation{Max Planck Institute for Solid State Research, Heisenbergstrasse 1, 70569 Stuttgart, Germany}
\author{Y. Liu}
\affiliation{Max Planck Institute for Solid State Research, Heisenbergstrasse 1, 70569 Stuttgart, Germany}
\author{K. S. Rabinovich}
\affiliation{Max Planck Institute for Solid State Research, Heisenbergstrasse 1, 70569 Stuttgart, Germany}
\author{G. Kim}
\affiliation{Max Planck Institute for Solid State Research, Heisenbergstrasse 1, 70569 Stuttgart, Germany}
\author{P. Wochner}
\affiliation{Max Planck Institute for Solid State Research, Heisenbergstrasse 1, 70569 Stuttgart, Germany}
\author{G. Christiani}
\affiliation{Max Planck Institute for Solid State Research, Heisenbergstrasse 1, 70569 Stuttgart, Germany}
\author{G. Logvenov}
\affiliation{Max Planck Institute for Solid State Research, Heisenbergstrasse 1, 70569 Stuttgart, Germany}
\author{K. Higuchi}
\affiliation{Max Planck Institute for Solid State Research, Heisenbergstrasse 1, 70569 Stuttgart, Germany}
\author{N. B. Brookes}
\affiliation{European Synchrotron Radiation Facility, Boîte Postale 220, F-38043 Grenoble, France}	
\author{E. Weschke}
\affiliation{Helmholtz-Zentrum Berlin für Materialien und Energie, BESSY II, Albert-Einstein-Str. 15, 12489 Berlin, Germany}	
\author{F. Yakhou-Harris}
\affiliation{European Synchrotron Radiation Facility, Boîte Postale 220, F-38043 Grenoble, France}
\author{A. V. Boris}
\affiliation{Max Planck Institute for Solid State Research, Heisenbergstrasse 1, 70569 Stuttgart, Germany}
\author{B. Keimer}
\affiliation{Max Planck Institute for Solid State Research, Heisenbergstrasse 1, 70569 Stuttgart, Germany}
\author{M. Minola}
\affiliation{Max Planck Institute for Solid State Research, Heisenbergstrasse 1, 70569 Stuttgart, Germany}

\begin{abstract}
Complex oxides are well known to develop oxygen ordering patterns with well defined periodicities, but their interplay with electronic correlations remains largely unexplored. Here, we report resonant and non-resonant x-ray diffraction data indicating a four-unit-cell periodic superstructure related to oxygen vacancy ordering in La$_{2-x}$Ca$_{x}$CuO$_{4}$ films with doping levels from the underdoped ($x = 0.15$) to the extremely overdoped ($x = 0.50$) regime. Whereas the lattice superstructure is temperature independent up to 300 K, a strongly temperature dependent electronic charge density wave (CDW) is observed in the underdoped and slightly overdoped regimes ($x\leq0.20$). The periodicity, in-plane and out-of-plane correlation lengths of the CDW are locked in by the lattice superstructure. Our results highlight the necessity to consider lattice and electronic energetics on equal footing in the high-temperature oxygen-disordered phase to explain oxygen ordering phenomena in complex oxides.

\end{abstract}
\maketitle

Atomic short-range order (ASRO) is ubiquitous in complex metal oxides. Partial order in the lattice site occupation by different ionic species has long been known to decisively influence the electronic properties of oxides \cite{Dagotto2005}, and is rapidly gaining prominence as a key performance characteristic of electrochemical devices \cite{Cai2024,Wang2024}. Whereas machine learning tools are increasingly applied to explore the thermodynamic stability of partially ordered metal oxides and to optimize their functionality \cite{Yang2021}, the microscopic mechanisms underlying the formation of ASRO in complex oxides remain largely unexplored. Theoretical research on a simpler set of materials -- disordered metallic alloys -- has offered detailed microscopic insight into the relationship between ASRO and the electronic structure \cite{Cowley1950,Moss1969,Ohshima1973,Gyorffy1983,Staunton1994,Wilkinson2001}. Specifically, the periodicity of ASRO is set by 
maxima of the electronic charge susceptibility, which naturally occur at wavevectors connecting parallel segments of the Fermi surface \cite{Moss1969}. Although band theory is typically associated with translationally symmetric lattices, it is now well-established that the notion of a Fermi surface remains relevant even in the high-temperature disordered phase of alloys, where dynamic short-range order is established as a precursor to ASRO \cite{Gyorffy1983,Staunton1994,Wilkinson2001}. 

Here we explore the relationship between ASRO and the electronic structure of the high-temperature superconducting cuprates, where ASRO has been widely studied in the form of partially ordered superstructures of oxygen ions \cite{Jorgensen1987,Beyers1989,Zimmermann2003,deFontaine1990,Anderson1999,Fontaine2005,Frello2000,Strempfer2004,Islam2004,Wells1996,Xiong1996,Lee2004,Frano2019,Izquierdo2011}. As oxygen ions act as dopants, their local correlations affect the local doping level and can induce spatial inhomogeneities in the electronic phase behavior. 
The discovery of short-range-ordered charge density waves (CDWs) with periodicities between 3 and 4 unit cells has recently opened a new perspective on the electronic structure of the cuprates. The CDW periodicity reflects features of the electronic charge susceptibility, which is strongly renormalized by Coulomb interactions between conduction electrons and difficult to compute in the strongly correlated electron system \cite{Comin2016,Frano2020,Hayden2024}.
CDWs have been widely observed in underdoped cuprates
 \cite{Hucker2011,Ghiringhelli2012,Tabis2014,Comin2014,Neto2016,Comin2016,Frano2020,Hayden2024,Betto2024}, and were more recently reported above the optimal doping level in certain systems: up to hole doping level of $p = 0.24$ in (Bi,Pb)$_{2.12}$Sr$_{1.88}$CuO$_{6+\delta}$ (Bi2201) \cite{Peng2018}, $p = 0.25$ in Tl$_2$Ba$_2$CuO$_{6+\delta}$ (Tl2201) \cite{Tam2022} and $p = 0.21$ in bulk La$_{2-x}$Sr$_x$CuO$_4$ (LSCO) \cite{Lin2020,Miao2021,vonArx2023}. 

Research on the relationship between ASRO and CDW order has thus far largely focused on the influence of the nonstatistical distribution of dopants on the CDW correlation length \cite{Achkar2014}. According to one recent report \cite{Li2023}, however, electronic CDW order with an unusual periodicity of $\sim 6$ unit cells is present in the extremely overdoped regime of LSCO with $p > 0.25$. Contrary to the underdoped systems, the purported CDW correlations are virtually temperature independent at least up to room temperature, raising the question to what extent lattice superstructures arising from cation or oxygen order may be involved at a more essential level.

\begin{figure}
	\includegraphics[width=0.47\textwidth]{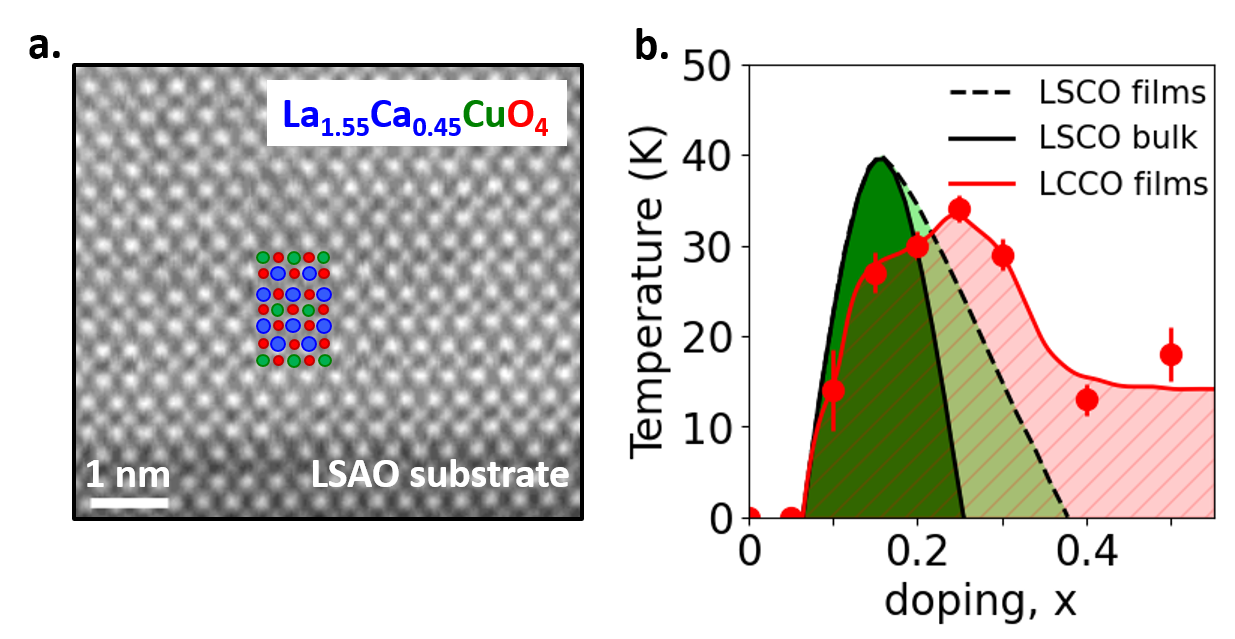}
	\caption{(a) Scanning transmission electron microscopy image of an $x = 0.45$ LCCO film, confirming high crystallinity. (b) Doping dependence of superconducting $T_c$ in 10 unit cell thick LCCO films compared with that of bulk LSCO and LSCO films \cite{Tallon1995,Dean2013,Sato2000,Kim2017}. Figures adapted from \cite{Kim2021}.}
	\label{fig1}
\end{figure}

Motivated by these reports, we have used a combination of Cu $L_3$-edge resonant x-ray scattering and non-resonant hard x-ray diffraction to investigate CDW correlations and lattice superstructures in thin films of La$_{2-x}$Ca$_x$CuO$_4$ (LCCO) over a wide range of hole-doping levels $x = 0.10 - 0.50$. LCCO is of particular interest because of its low level of structural disorder due to the nearly identical radii of La and Ca ions. Our films were extensively characterized in prior work \cite{Kim2021} (see also Fig.~\ref{fig1}(a) and \cite{SM}), and it was found that  superconductivity in this system persists to a doping level of at least $x = 0.50$, well beyond what is known for systems with higher structural disorder such as LSCO (Fig.~\ref{fig1}(b)).
Here we report a temperature independent bulk four-unit-cell periodic lattice superstructure over an exceptionally wide doping range, $x = 0.15 - 0.50$, which we attribute to oxygen-vacancy ordering. For $x = 0.15$ and 0.20, we additionally find a strongly temperature dependent electronic CDW with periodicity, in-plane and out-of-plane correlations locked into those of the lattice superstructure. Our results indicate that lattice and electronic instabilities are highly interdependent, highlighting the need to treat both phenomena self-consistently. At low temperatures, the static oxygen superstructure clearly serves as a template for the electronic CDW. Conversely,  the unusual stability of the four-unit-cell oxygen superstructure and its coincidence with the CDW raise the question how electronic correlations contribute to the formation of dynamical oxygen short-range order at high temperatures, where oxygen diffusion kinetics is appreciable, and hence to the static oxygen superstructure observed in our low-temperature experiments. In analogy to ASRO in metallic alloys, our results call for high-temperature electronic-structure calculations to address this issue. 
	
LCCO films of 10 and 100 unit cell thickness ($\sim 13.2$ nm and 132 nm, respectively) were grown on LaSrAlO$_4$ (LSAO) (001) substrates using molecular beam epitaxy and extensively characterized, as reported previously \cite{Kim2021} (see also Fig.~\ref{fig1}(a,b) and \cite{SM}). The resonant inelastic x-ray scattering (RIXS) measurements were performed using the ERIXS spectrometer at the ID32 beamline at the European Synchrotron Radiation Facility in Grenoble. The scattering angle was fixed at 149.5$^\text{o}$. Since our focus is on static CDW and lattice superstructures, the RIXS measurements were performed with an energy resolution of 45 meV, and we discuss only the resolution-limited (quasi)elastic scattering part of the RIXS spectra (see \cite{SM} for examples of full spectra). The resonant energy-integrated x-ray scattering (REXS) measurements were performed at the beamline UE46-PGM1 at the BESSY-II synchrotron of the Helmholtz-Zentrum in Berlin. All resonant x-ray measurements were performed at the Cu $L_3$ absorption edge ($\sim$ 931.5 eV) with $\sigma$ incident polarization, unless otherwise specified. The non-resonant hard x-ray diffraction measurements were performed with an incident x-ray energy of 10 keV at the MPI beamline of the KARA synchrotron. The resonant x-ray and hard x-ray measurements were performed on 10 unit cell and 100 unit cell thick films, respectively.
	
\begin{figure}
	\includegraphics[width=0.49\textwidth]{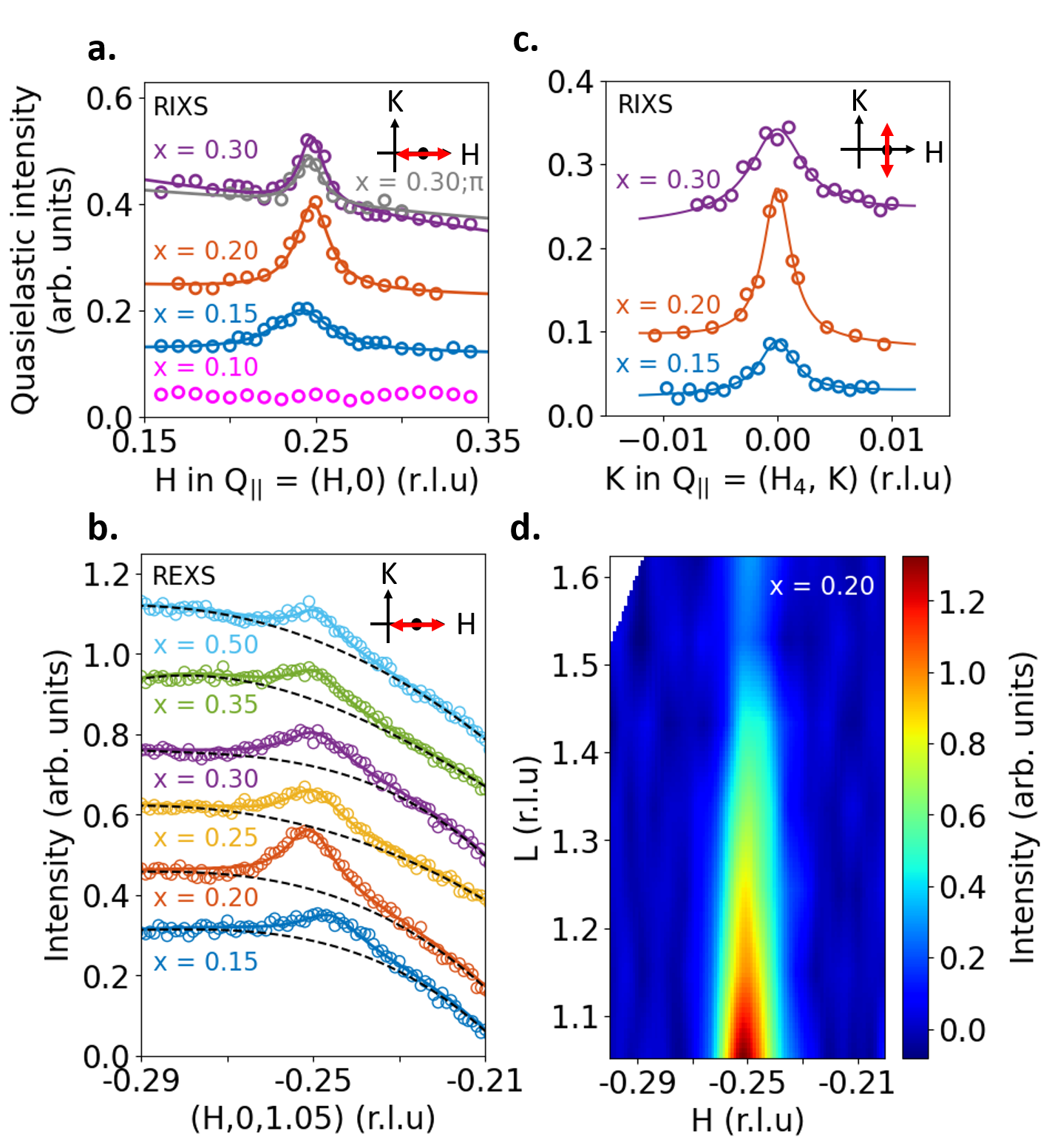}
	\caption{(a) $H$ scan of the energy-integrated quasielastic scattering intensity obtained from RIXS for various dopings. (b) $H$ scan of the REXS intensity at $L = 1.05$, obtained for various dopings. The dashed black lines represent the fitted third-order polynomial background. (c) $K$ scan of the energy-integrated quasielastic scattering intensity obtained from RIXS across $Q_4 = (H_{4},0)$. The solid lines in (a-c) represent the results of fits to the data as described in the text. (d) Reciprocal space map of the background-subtracted REXS intensity in the $H-L$ plane for the $x = 0.20$ sample. The data were linearly interpolated to create the colormap. All data were obtained at $T = T_c$.}
	\label{fig2}
\end{figure}

We first present data from Cu $L_3$ edge RIXS, which are highly sensitive to charge correlations of the Cu $3d$ conduction electrons. Fig.~\ref{fig2}(a) displays the momentum dependence of the energy-integrated quasi-elastic scattering intensity (integration range -0.15 eV to 0.1 eV) along the copper-oxide bond direction ($H$) at various doping levels. Note that the momentum coordinates are given in reciprocal lattice units, so that $H_4 = 1/4$ corresponds to four unit cells in real space. Clear peaks near $H_4$ are observed for the dopings $x = 0.15,$ 0.20, 0.25 and 0.30, whereas no peak is observed for $x = 0.10$. Upon switching the incident beam to $\pi$ polarization, the peak in the $x = 0.30$ sample becomes weaker, confirming that the observed quasi-elastic scattering signal originates from charge correlations \cite{Ghiringhelli2012}. 
Further confirmation of the four-unit-cell reconstruction was obtained from REXS measurements as shown in Fig.~\ref{fig2}(b). The highest peak amplitude is observed for $x = 0.20$, in agreement with the RIXS results. Due to the advantage of faster measurements possible with REXS, more dopings could be explored, and analogous peaks were  also observed in samples with doping levels of $x = 0.35$ and 0.50. Using least-squares fits to Lorentzian profiles for the peak and a first (third) order polynomial for the smooth background in RIXS (REXS), we extracted the characteristic wavevector, which evolves from $H\sim 0.245$ for $x = 0.15$ to $\sim 0.25$  for all dopings $x \geq 0.20$. An equivalent peak was observed in the $(0,K)$ direction (see \cite{SM}), as expected based on the tetragonal symmetry of the crystal structure. For simplicity, we refer to both wavevectors as $Q_4$. The correlation length parallel to $Q_4$, $\xi_{||} \sim 60$ \r{A}, extracted from the fits is typical of electronic CDWs in cuprates \cite{Frano2020}. The in-plane correlation length perpendicular to $Q_4$ was extracted from scans parallel to the $K$ ($H$) direction across $(H_4,0)$ [$(0,K_4)]$. The value obtained in this way, $\xi_{\perp} \sim 200 - 300$ \r{A},  is  about 4-5 times larger than $\xi_{||}$ (see \cite{SM} and Fig.~\ref{fig2}(c)), indicating the uniaxial nature of the corresponding domains \cite{HHKim2021}.  

\begin{figure}
	\includegraphics[width=0.48\textwidth]{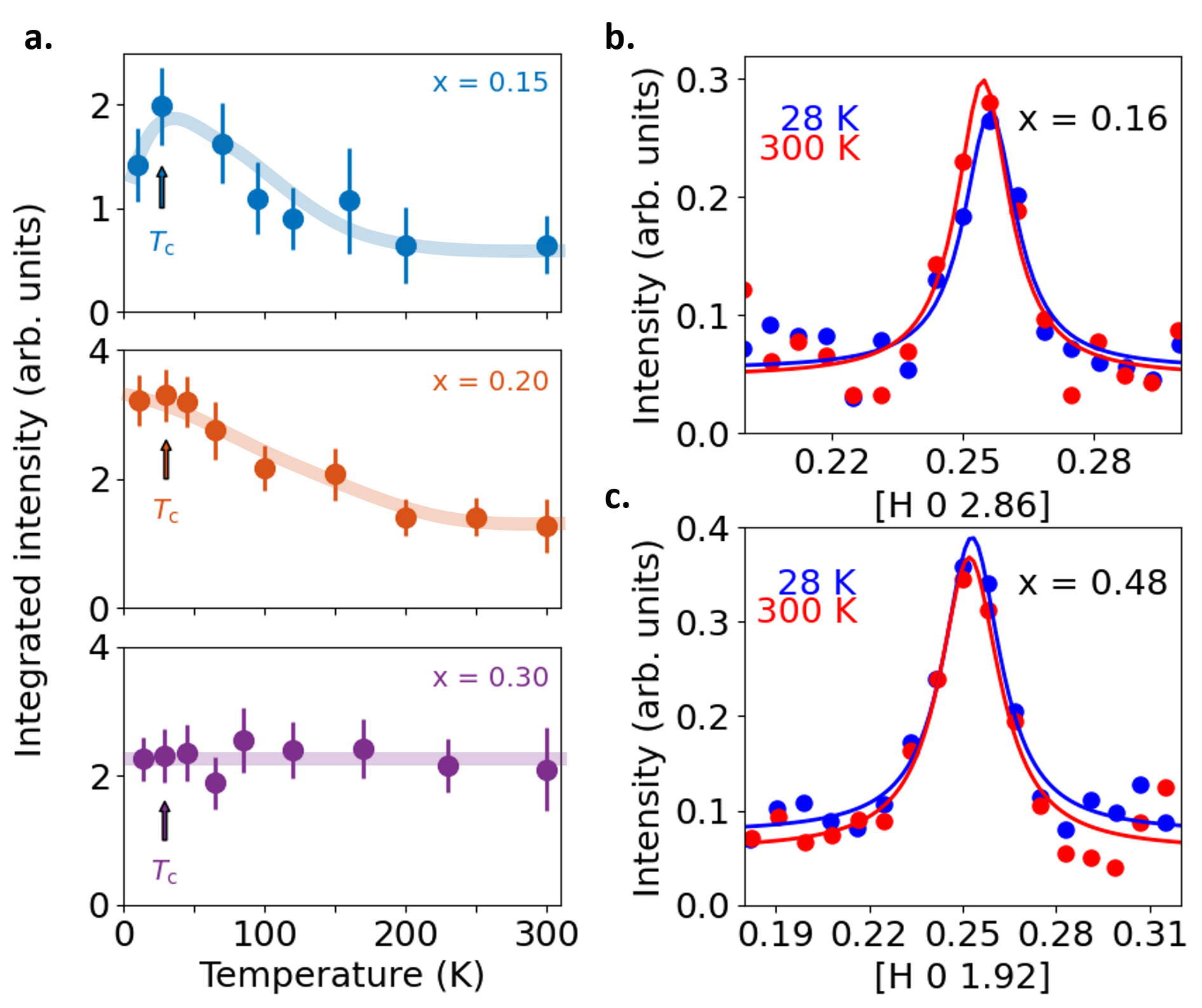}
	\caption{(a) Temperature dependence of the integrated intensity of the charge order peak obtained from REXS for $x = 0.15,$ 0.20 and 0.30. The $T_c$ of each sample is indicated. (b,c) $H$ scans across $Q_4 = (H_{4},0)$ obtained for 100 unit cell thick $x = 0.16$ and $x = 0.48$ samples, respectively, at 28 K and 300 K, with non-resonant hard x-ray diffraction. The solid lines were obtained from Lorentzian fits.}
	\label{fig3}
\end{figure}

Next, we look at the correlations of the superstructure signal along the $c$-axis, by mapping out its $L$-dependence using REXS (see Fig.~\ref{fig2}(d) and \cite{SM}). Although the $Q$-range accessible with Cu $L_3$-edge photons is limited to $L \geq 1.05$, the salient features of the out-of-plane correlations are apparent from the resulting data. In particular, the peak intensity increases upon approaching  $L \sim 1$ over the entire doping range $x = 0.15 - 0.50$, unlike previous observations of CDWs in bulk single crystals of LSCO and La$_{2-x}$Ba$_x$CuO$_4$ (LBCO), where the strongest out-of-plane CDW correlations appear at half-integer $L$ \cite{Hucker2011,Croft2014,Miao2021}. (Note that the $L$-dependence measurements were not attempted with RIXS due to their time-consuming nature.)

Since electronic CDWs are generally expected to display a strong temperature dependence, determination of the temperature dependence of the superstructure peaks is a possible diagnostic of their electronic \textit{vs.} structural nature. Figure~\ref{fig3}(a) displays the $T$-dependence of the integrated peak intensity for $x = 0.15, 0.20$ and 0.30 obtained using REXS. A strong temperature dependence and onset-like behavior at $T_{\text{onset}}$ $\sim 150 - 200$ K is observed for $x = 0.15$ and $x = 0.20$. A downturn of the $Q_4$ intensity below $T_c$ can be discerned for $x = 0.15$ (albeit within the error bar), evidencing competition between the CDW and superconductivity. These observations are consistent with prior reports of electronic CDW in bulk single crystals \cite{Tranquada1995,Hucker2011,Thampy2014,Croft2014,Wen2019,Miao2021}. On the other hand, the intensity is nearly temperature independent for $x = 0.30$. Limited $T$-dependence data collected on $x = 0.35$ (see \cite{SM}) also show nearly temperature independent behavior. 



Several studies on bulk LSCO have revealed nearly temperature independent CDW signals in certain regions of the temperature-doping phase diagram, which were attributed to short range CDW order \cite{Miao2021,vonArx2023}. Recent studies on YBCO have also pointed out that a charge fluctuation regime persists in an extended region of the phase diagram, displaying similar temperature independent behavior \cite{Arpaia2019,Arpaia2023}. Such fluctuations are significantly shorter-ranged than the static CDW. However, the correlation lengths we extract from our measurements are nearly temperature- and doping-independent with $\xi_{||}$ remaining at $\sim 60-80$ \r{A} (see \cite{SM}), implying that the $T$-independent component cannot be attributed to short range CDW order. Instead, the crossover in the temperature dependence of the integrated intensity observed in our LCCO films with increasing doping strongly suggests that the peak at higher dopings ($x \geq 0.30$) has a purely structural rather than an electronic origin, whereas the peak at lower dopings ($x \leq 0.20$) displays signatures of a partially electronic origin common to CDWs in other cuprates. 


Further confirmation of the structural nature of the $T$-independent intensity comes from non-resonant hard x-ray diffraction measurements. Whereas the resonance at the Cu $L_3$ edge greatly enhances the contribution of Cu $3d$ electronic charge modulations to the total diffraction signal, non-resonant measurements are sensitive to the total electron density and are hence dominated by lattice contributions. Figure~\ref{fig3}(b,c) shows $H$ scans across $Q_4$ obtained for films with $x = 0.16$ and $0.48$, respectively, using non-resonant hard x-ray diffraction. Clear temperature independent peaks are observed near $H = 0.25$ for both samples. This observation demonstrates that the temperature-independent signal observed for $x \geq 0.30$ and the temperature-independent high-$T$ intensity observed for $x = 0.15$ and 0.20 in Fig.~\ref{fig3}(a) are associated with a lattice superstructure. 

\begin{figure}
	\includegraphics[width=0.42\textwidth]{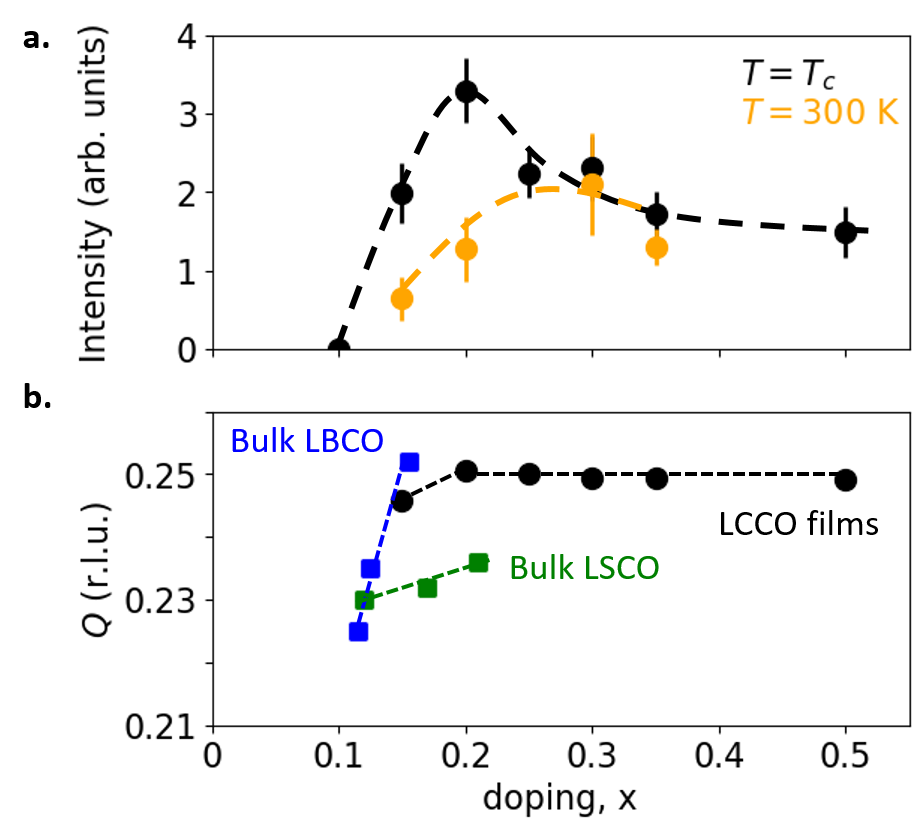}
	\caption{Doping dependence of (a) the integrated intensity of the superstructure peak, and (b) its characteristic wavevector $Q$, extracted from fits to the REXS data at $T = T_c$. The integrated intensity at 300 K and the CDW wavevectors for bulk LSCO \cite{Croft2014,Miao2021} and LBCO \cite{Miao2019} are included in (a,b) respectively for comparison. The dashed lines are guides to the eye.}
	\label{fig4}
\end{figure}

What is the origin of the four-unit-cell lattice superstructure? Tilt distortions of the oxygen octahedra are known to cause structural phase transitions in LSCO and LBCO \cite{Axe1994,Pelc2022}. However, such distortions produce only half-integer superstructural reflections and disappear around $x \sim 0.25$ \cite{Axe1994,Pelc2022}. Furthermore, LCCO is expected to be less prone to such distortions due to the similar ionic radii of La and Ca. In addition, the structure of the film is locked into that of the tetragonal LSAO substrate, making such distortions less likely in films. Two other possibilities arise, namely Ca dopant ordering and oxygen ordering. If Ca dopant ordering were responsible, one would expect a strong dependence of the intensity, periodicity and correlation length of the superstructure on the Ca doping level, which is not observed in our measurements (see Fig.~\ref{fig4}(a,b) and \cite{SM}). On the other hand, previous \textit{in situ} RHEED measurements have documented a reversible switching between 4- and 5-unit-cell configurations of the surface structure during growth under vacuum \textit{vs.} ozone-rich atmosphere, respectively \cite{Suyolcu2021} (see also \cite{SM}). Furthermore, prior Raman spectroscopy studies on LSCO films have shown that the overdoped regime is particularly prone to apical oxygen vacancy formation \cite{Kim2017}. These observations indicate that the superstructure we detect is most likely associated with oxygen vacancy ordering.

Optical spectroscopy measurements on LCCO films, in conjunction with electronic-structure calculations, indicate a progressively increasing density of holes in the Cu 3$d_{3z^2-r^2}$ orbital with increasing doping for $x \geq 0.15$ \cite{Weber2010,Kim2021}. We cannot rule out a charge-and-orbital ordering scenario with full site occupany at the apical-oxygen position. In this case, however, we would expect a systematic increase of the REXS signal with increasing $x$, which is not observed in our measurements (see Fig.~\ref{fig4}(a)). 

From the considerations above, we infer that the superstructure peak at doping levels $x \leq 0.20$ is a superposition of the temperature independent oxygen superstructure and a temperature dependent electronic CDW templated by this structure. We note that locking of electronic charge-order correlations by the crystal structure was also recently reported in LBCO, where the formation of the low-temperature tetragonal phase coincides with a sudden jump and lock-in of an electronic charge-order reflection \cite{Miao2019}. To determine the electronic CDW component of superstructure peaks observed with resonant x-ray scattering, one therefore needs to look at the contrast between low-$T$ and high-$T$ integrated intensities. Figure~\ref{fig4}(a) shows the doping dependence of the integrated intensity of the peaks obtained at $T = T_c$ and 300 K. Indeed, a significant difference between the intensities obtained at the two temperatures exists only for $x = 0.15$ and 0.20. This confirms that substantial electronic CDW correlations are present only in this doping range, in agreement with prior reports on bulk LSCO \cite{Lin2020,Miao2021}. The increase in the characteristic CDW wavevector with doping agrees with prior CDW reports on bulk LSCO \cite{Croft2014,Miao2021} and LBCO \cite{Miao2019} (see Fig.~\ref{fig4}(b)) . The stronger CDW intensity at $x = 0.20$, compared to $x = 0.15$, may reflect the van-Hove singularity in the electronic density of states at $x = 0.19$. \cite{Yoshida2006,Horio2018,Miao2021}

We now discuss our results in the context of recent experiments on different systems. In particular, a recent study of LSCO films reported superstructure reflections with $Q_6 \sim 1/6$ in the extremely overdoped regime ($x = 0.35, 0.45$ and 0.60) \cite{Li2023} and assigned those reflections to an electronic CDW. If this assignment were correct, this would imply a major revision of the conventional electronic phase diagram of the cuprates, which classifies highly overdoped systems as featureless Fermi liquids. We therefore searched for peaks in a similar $Q$ range in our LCCO films, with negative results (see \cite{SM}). We note, however, that the reflections reported in Ref. \onlinecite{Li2023} are almost $T$-independent, as the $Q_4$ reflections we observed in LCCO, and that they display a closely similar in-plane correlation length and $L$-dependence. These observations point to a lattice superstructure -- likely a different oxygen ordering pattern -- rather than an electronic CDW.

A four-unit-cell oxygen superstructure has also been reported for optimally doped YBCO \cite{Strempfer2004,Islam2004}, suggesting that the mechanisms underlying this phenomenon may be generic to the cuprates. Prior scanning tunneling microscopy studies reported a doping independent four-unit-cell periodic charge modulation at the surface of Bi$_2$Sr$_2$CaCu$_2$O$_{8+\delta}$ (Bi2212) over a wide doping range, which was interpreted as a commensurate CDW \cite{Mesaros2016}. 
However, these measurements were limited to low temperatures, and could therefore not conclusively differentiate between electronic and (possibly sub-surface) oxygen superstructures, necessitating additional measurements with complementary techniques. Our study demonstrates the importance of determining the $T$-dependence to discriminate between electronic and lattice contributions to superstructures in the cuprates. 

It is also worth noting that RIXS and REXS experiments on films of the infinite-layer nickelates NdNiO$_2$ and PrNiO$_2$ uncovered a superstructure reflection at $Q_3 \sim (1/3,0)$ with weakly temperature dependent intensity and correlation length \cite{rossi2022,krieger2022,ren2023}, similar to our observations for LCCO films with $x > 0.20$. Whereas the reflection was originally attributed to a CDW, follow-up work indicates that the observed signal is at least partially associated with oxygen ordering in insufficiently reduced films \cite{parzyck2024,pelliciari2023comment} (see also \cite{Hayashida2024}).

In conclusion, our data show that the four-unit-cell oxygen superstructure in LCCO is stable over an exceptionally wide range of Ca concentrations ($x = 0.15 - 0.50$). Together with related observations in other cuprate families, our results indicate that this superstructure may be rooted in interactions that are generic to the cuprates. The locking of the electronic CDW by oxygen vacancy order that we have observed in LCCO underscores the interdependence of both phenomena. Clearly, the oxygen superstructure serves as a template for spatial correlations amongst the valence electrons -- but could electronic correlations in turn influence positional correlations between oxygen ions? 

In the cuprates, these correlations and the resulting stability of different oxygen-ordered phases have thus far only been described in terms of semi-empirical parameters \cite{deFontaine1990,Anderson1999,Fontaine2005}. By treating the lattice energetics and electronic structure self-consistently, research on disordered metallic alloys has offered microscopic insight into the formation of ASRO, which is also instructive for the case at hand \cite{Cowley1950,Moss1969,Ohshima1973,Gyorffy1983,Staunton1994,Wilkinson2001}. Specifically, the oxygen ions are mobile during synthesis of oxide compounds at high temperatures, and their positional correlations are expected to be influenced by the charge susceptibility of the conduction electron gas  -- the high-temperature remnant of the susceptibility that generates the CDW order observed in our low-temperature RIXS and REXS experiments. The oxygen superstructure that is frozen in upon cooling in turn affects the periodicity and correlation volume of the CDW \cite{Vinograd2024}. Ab-initio electronic structure theory and thermodynamics will be required to fully elucidate the interdependence of ASRO and electronic correlations in cuprates, nickelates, and other complex oxides. Control of ASRO during growth \cite{Suyolcu2021} could serve as an additional tuning knob to develop oxide materials for enhanced functional properties. As the interplay between ASRO and such properties is also of growing importance for electrochemical devices \cite{Wang2024,Cai2024,Yang2021}, this line of research will open a new frontier at the nexus of quantum and energy science.

The RIXS data were collected at the beam line ID32 of the European Synchrotron Radiation Facility (ESRF) in Grenoble, France (proposal \# 5041). The REXS data were collected at the beamline UE46-PGM1 of the Helmholtz-Zentrum Berlin at BESSY-II, Berlin. The Institute for Beam Physics and Technology (IBPT) at the Karlsruhe Institute for Technology
(KIT) is acknowledged for the operation of the storage ring, the Karlsruhe Research Accelerator (KARA), and provision of beamtime at the KIT light source. S. H. is supported by the Alexander von Humboldt Foundation.

\bibliography{LCCO_ChargeOrder.bib}

\widetext
\clearpage

\begin{center}
	\textbf{\large Supplemental Material}
\end{center}

Here we document XRD characterization of the $x = 0.15$ sample (Fig.~\ref{fig:RSM_x15}), \textit{in situ} RHEED characterization of the $x = 0.50$ sample (Fig.~\ref{fig:RHEED}), examples of raw RIXS spectra (Fig.~\ref{fig:RIXS_raw}), evidence for presence of the superstructure parallel to the second tetragonal axis (Fig.~\ref{fig:KscanHcross}), REXS scans across a wide momentum range (Fig.~\ref{fig:REXS_extended}), raw data for the $L$ dependence (Fig.~\ref{fig:Ldepx20} and Fig.~\ref{fig:Ldepx15to35}) and temperature dependence (Fig.~\ref{fig:Tdepx15to30}) of the superstructure peak obtained from REXS, the temperature dependence of in-plane correlation length (Fig.~\ref{fig:Tdep_CorrLength_x15to30}) and $Q$ (Fig.~\ref{fig:Tdep_Qco_x15to30}) of the superstructure peak.

\clearpage
\setcounter{equation}{0}
\setcounter{figure}{0}
\setcounter{table}{0}
\setcounter{page}{1}
\makeatletter
\renewcommand{\theequation}{S\arabic{equation}}
\renewcommand{\thetable}{S\arabic{equation}}
\renewcommand{\thefigure}{S\arabic{figure}}
\renewcommand{\thesection}{S\arabic{section}}
\renewcommand{\bibnumfmt}[1]{[S#1]}

\section{XRD and RHEED characterization}

\begin{figure}[!htb]
	\includegraphics[width=\textwidth]{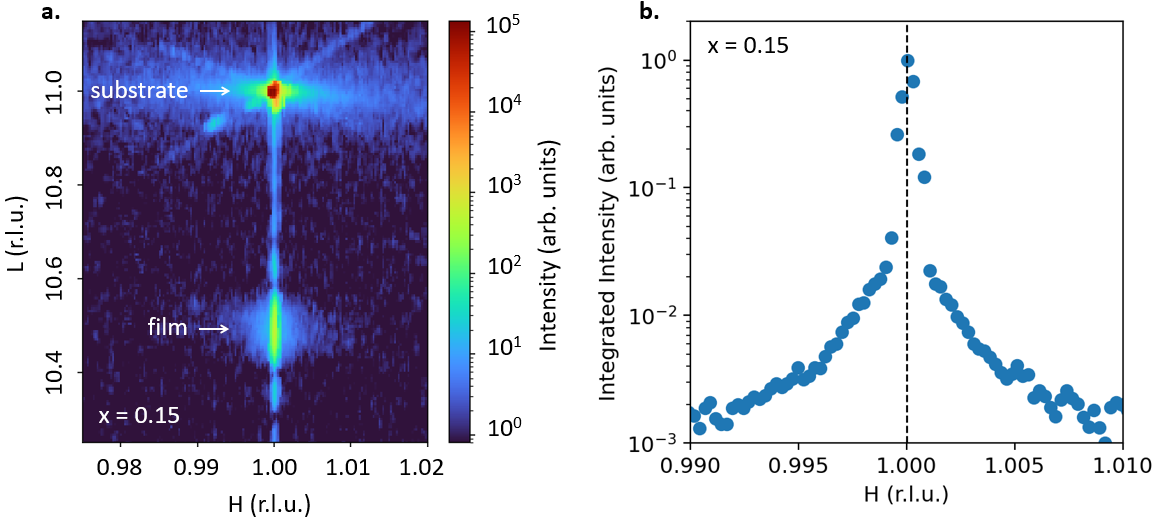}
	\caption{(a,b) Reciprocal space map (a) and corresponding $L$-integrated intensity \textit{vs.} $H$ (b) obtained for the 10 unit cell thick $x = 0.15$ sample using a Bruker D8 Discover XRD machine equipped with a Cu K$\alpha$ source. Note that the units used here for $H$ and $L$ correspond to the reciprocal lattice units of the substrate. The film peak is observed to appear at the same $H$ value as that of the substrate peak, confirming that the film is fully strained. }
	\label{fig:RSM_x15}
\end{figure}

\begin{figure}[!htb]
	\includegraphics[width=0.4\textwidth]{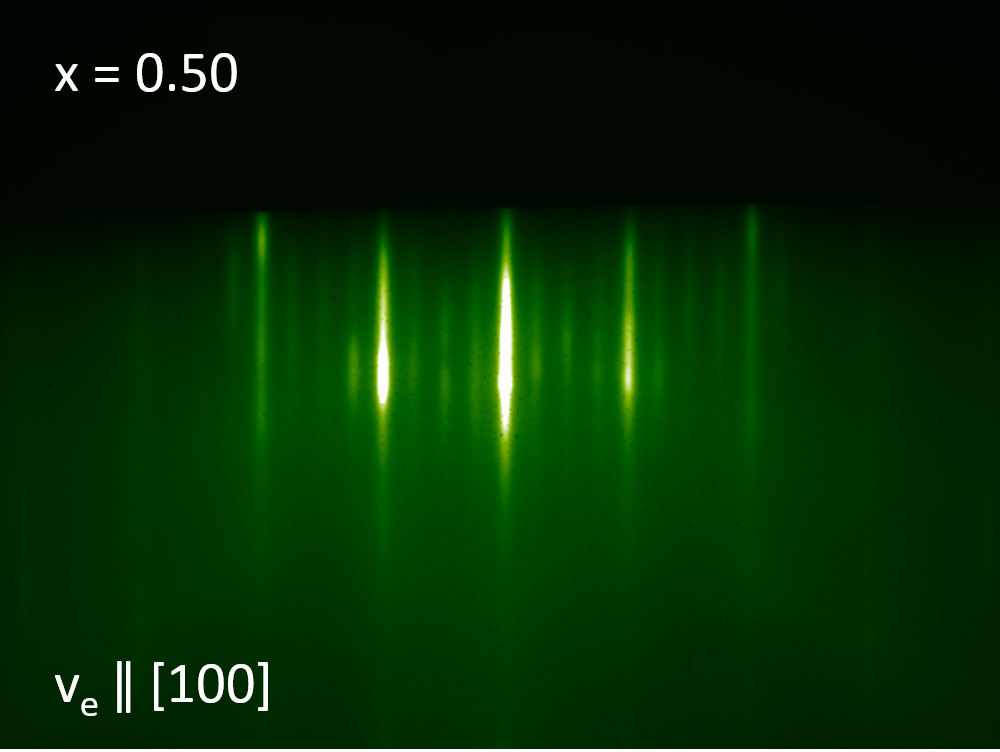}
	\caption{In-situ RHEED pattern of an $x = 0.50$ LCCO film. The incoming electron beam was directed along [100]. Three weak satellite streaks are observed between the main streaks, indicating a four-unit-cell periodic surface superstructure.}
	\label{fig:RHEED}
\end{figure}

\clearpage

\section{RIXS - additional data}
\label{RIXS}

\begin{figure}[!htb]
	\includegraphics[width=0.55\textwidth]{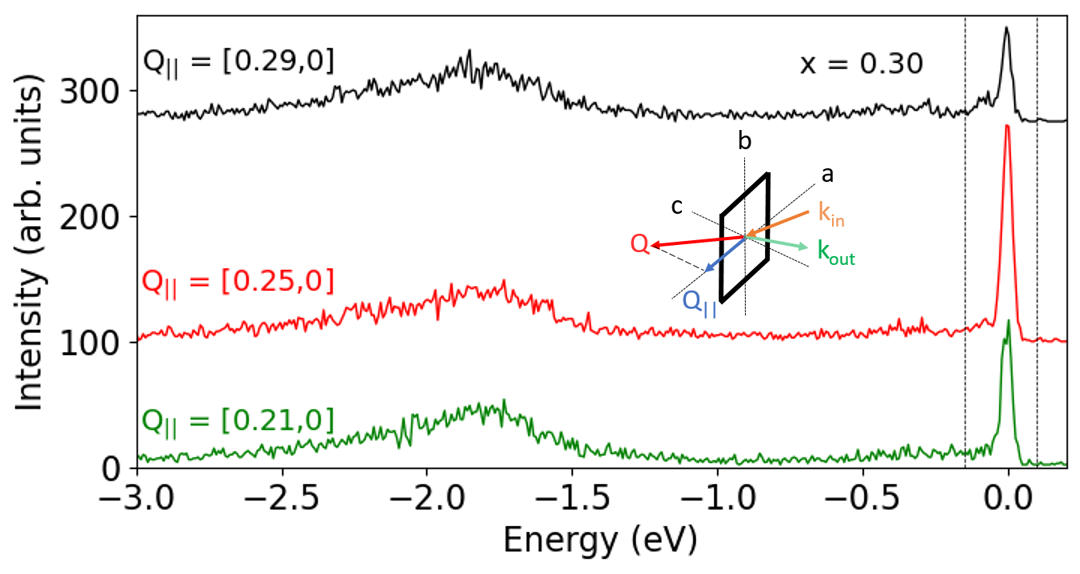}
	\caption{Raw RIXS spectra obtained for $x = 0.30$ at different in-plane momentum vectors $Q_{||}$. The dashed lines indicate the energy integration range used to extract the quasielastic scattering intensity. The inset shows the RIXS and REXS scattering geometry.}
	\label{fig:RIXS_raw}
\end{figure}

\begin{figure}[!htb]
	\includegraphics[width=0.7\textwidth]{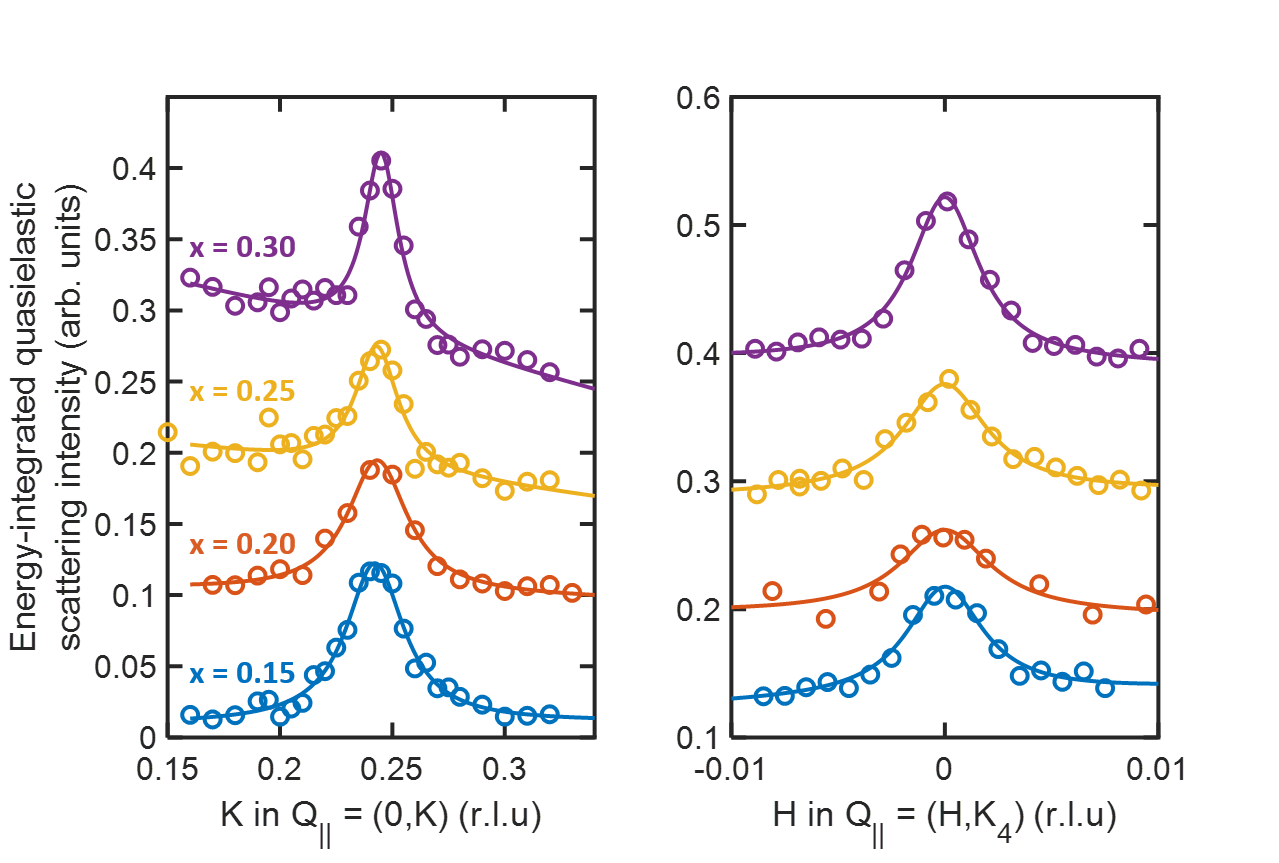}
	\caption{(a) $K$ scan of the energy-integrated quasielastic scattering intensity obtained from RIXS for various dopings. (b) $H$ scan of the energy-integrated quasielastic scattering intensity obtained from RIXS across the characteristic wavevector (0,$K_{4}$) of the peak in (a). The solid lines represent fits to the data as described in the main text. All data were obtained at $T = T_c$.}
	\label{fig:KscanHcross}
\end{figure}

\clearpage
\section{REXS - additional data}

\begin{figure}[!htb]
	\includegraphics[width=0.35\textwidth]{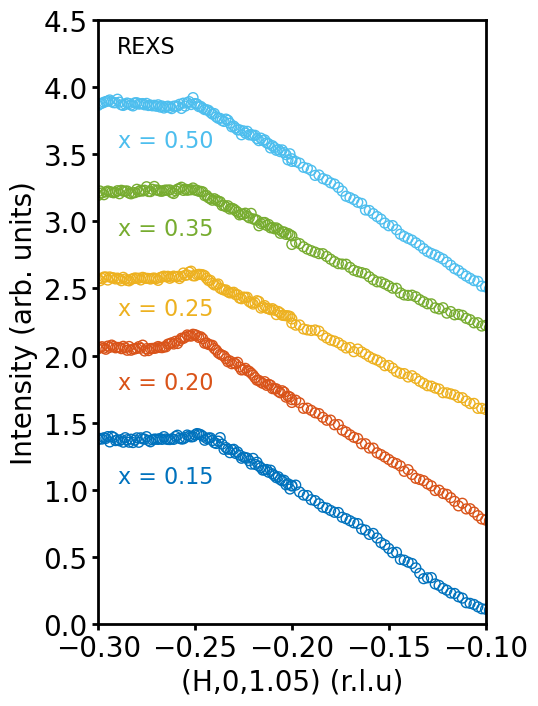}
	\caption{Wide range $H$ scan of the REXS intensity at $L = 1.05$, obtained for various dopings. The $H$ range includes $H_6 = 1/6$ where CDW was recently reported in extremely overdoped LSCO films \cite{Li2023}. }
	\label{fig:REXS_extended}
\end{figure}

\begin{figure}[!htb]
	\includegraphics[width=0.35\textwidth]{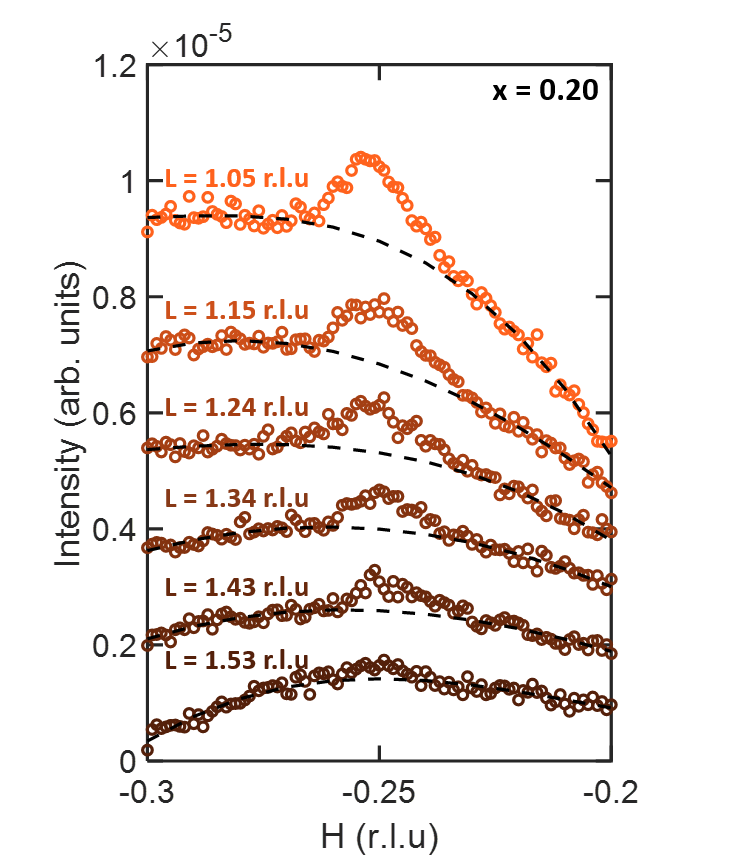}
	\caption{Raw $H$ scans of the REXS intensity at different fixed $L$ obtained for $x = 0.20$. The dashed black lines represent the fitted third-order polynomial background.}
	\label{fig:Ldepx20}
\end{figure}

\begin{figure}[!htb]
	\includegraphics[width=\textwidth]{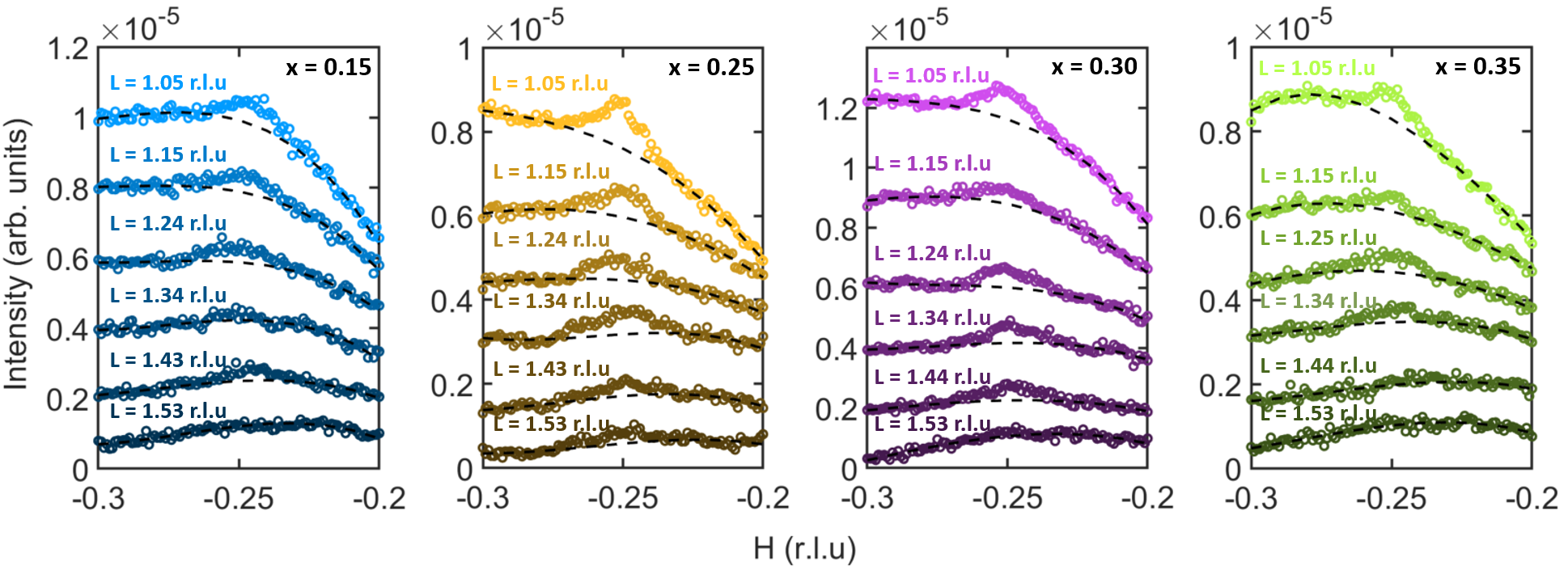}
	\caption{Raw $H$ scans of the REXS intensity at different fixed $L$ obtained for $x = 0.15, 0.25, 0.30$ and 0.35. The dashed black lines represent the fitted third-order polynomial background.}
	\label{fig:Ldepx15to35}
\end{figure}

\begin{figure}[!htb]
	\includegraphics[width=\textwidth]{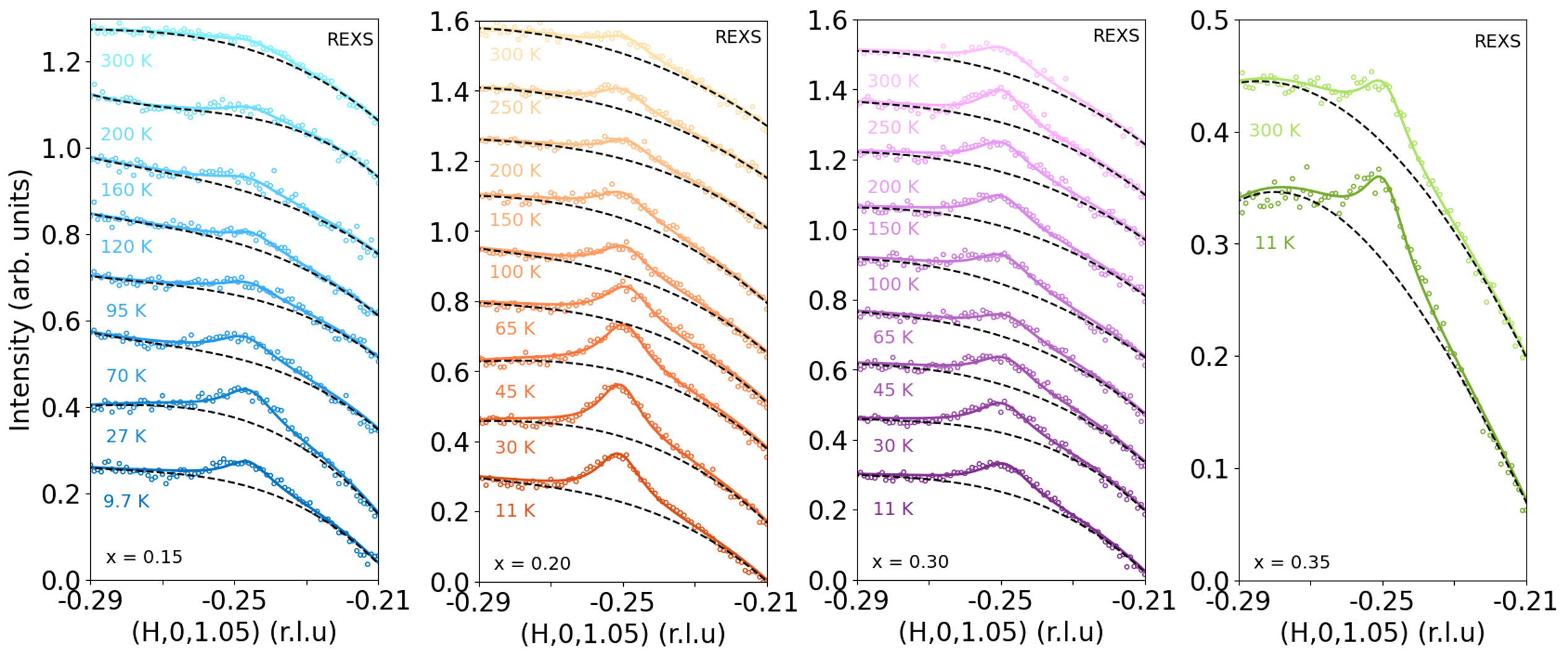}
	\caption{Raw $H$ scans of the REXS intensity at $L = 1.05$ obtained for $x = 0.15, 0.20, 0.30$ and 0.35 at different temperatures. The dashed black lines represent the fitted third-order polynomial background. The solid lines represent fits to the peak as described in the main text.}
	\label{fig:Tdepx15to30}
\end{figure}

\begin{figure}[!htb]
	\includegraphics[width=0.7\textwidth]{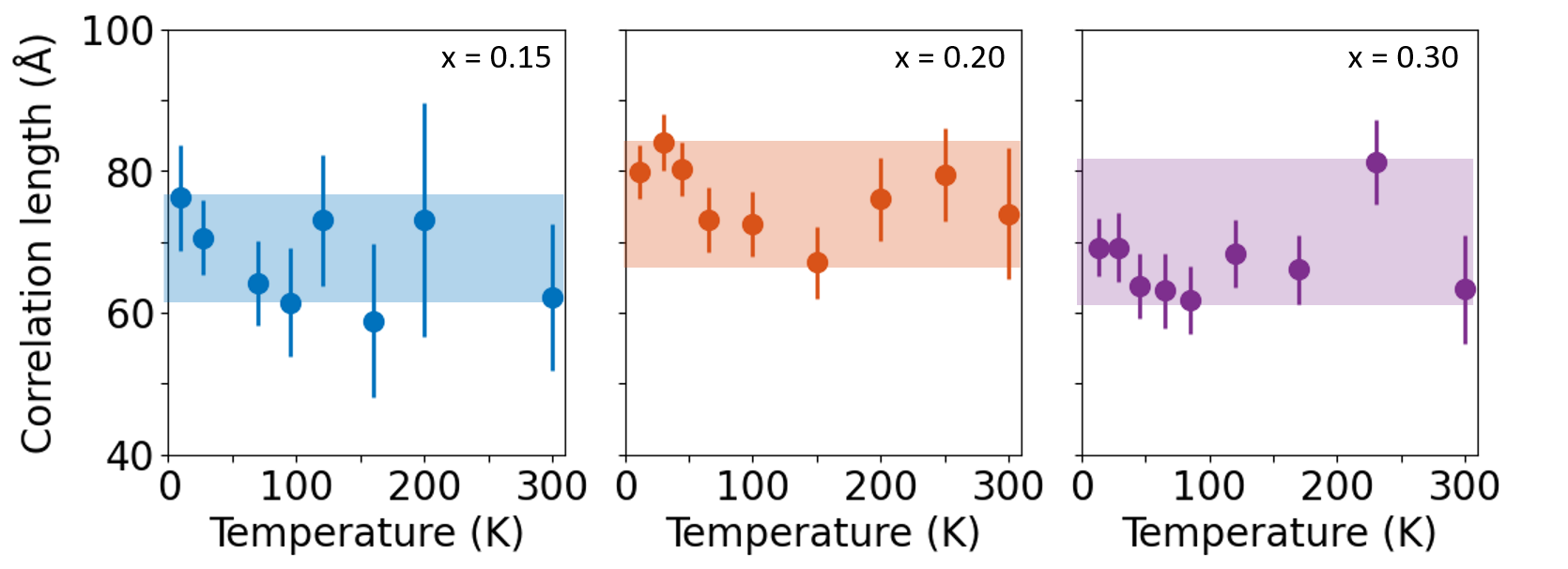}
	\caption{Temperature dependence of the longitudinal correlation length for $x = 0.15, 0.20$ and 0.30, obtained from the fits in Fig.~\ref{fig:Tdepx15to30}.}
	\label{fig:Tdep_CorrLength_x15to30}
\end{figure}

\begin{figure}[!htb]
	\includegraphics[width=0.7\textwidth]{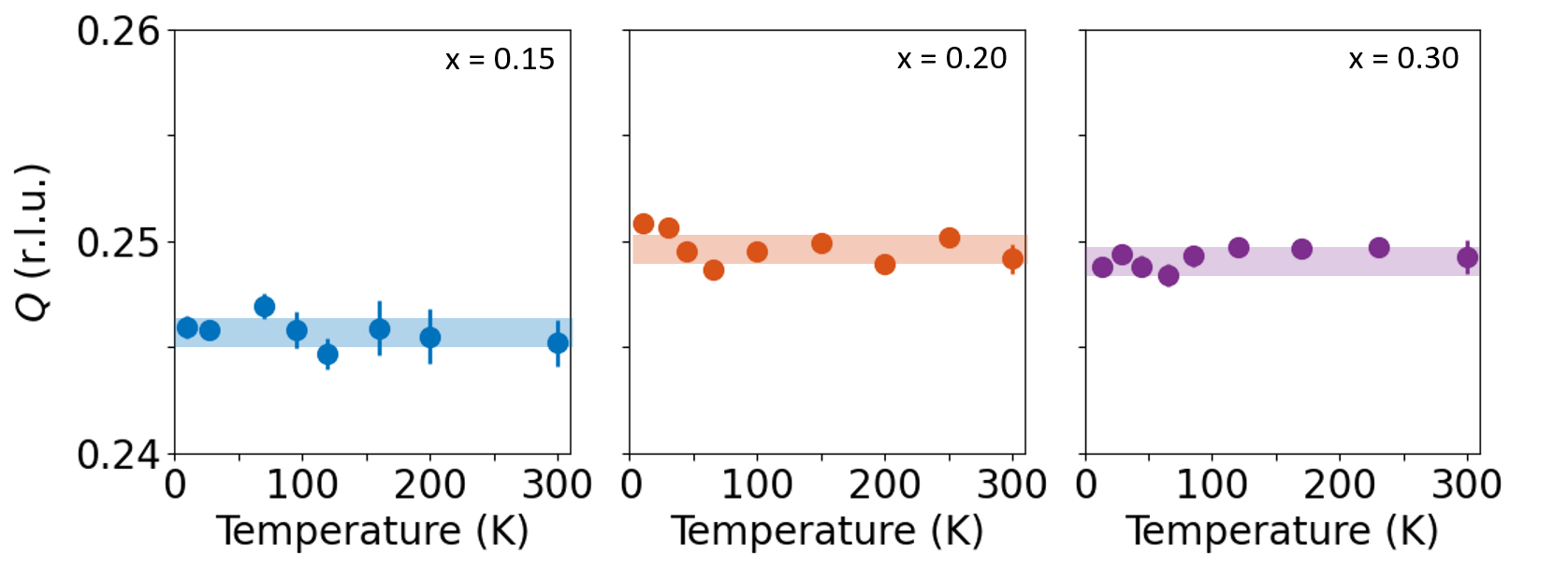}
	\caption{Temperature dependence of the wavevector characterizing the superstructure peak in $x = 0.15, 0.20$ and 0.30, obtained from the fits in Fig.~\ref{fig:Tdepx15to30}.}
	\label{fig:Tdep_Qco_x15to30}
\end{figure}

\end{document}